\documentclass[english,prl,reprint,showpacs,showkeys]{revtex4-1}
\usepackage[T1]{fontenc}
\usepackage[utf8]{inputenc}
\usepackage{textcomp}
\usepackage{amsmath}
\usepackage{amssymb}
\usepackage{esint}

\makeatletter

\providecommand{\tabularnewline}{\\}

\usepackage{slashed}
\usepackage{bbm}

\makeatother

\usepackage{babel}
\begin{document}

\title{On the nature of the neutrino}

\author{R. Romero}

\email{ricardo.romero@xanum.uam.mx}

\affiliation{Departamento de Física, División de Ciencias Básicas e Ingeniería,
Universidad Autónoma Metropolitana unidad Iztapalapa, San Rafael Atlixco
N\textdegree{} 186, Col. Vicentina, C.P. 09340, Del. Iztapalapa, México
D.F. }
\begin{abstract}
Assuming that one neutrino type with definite mass is described by
a massive Dirac field operator, it is shown that the physical one-particle
states for particles and antiparticles can be rotated to each other,
irrespective of their helicity. This result is used to prove that
the neutrino must necessarily be a Majorana particle. 
\end{abstract}

\pacs{14.60.Lm, 11.10.-z, 11.30.Er}

\keywords{Dirac neutrino, Majorana neutrino, momentum-dependent rotations.}

\maketitle

\section{Introduction}

Originally conceived to be massless in the standard model, neutrinos
are now known to oscillate in flavor and posses tiny but non-vanishing
masses \cite{Agashe:2014kda,Bilenky:2014ema,doi:10.1142/S0217751X0904703X},
although their absolute values are unknown, with data and estimations
about squared-mass splittings and upper bounds for mass eigensates
currently available \cite{Agashe:2014kda}. Disregarding gravity,
neutrinos interact only through the weak interaction, and only left-handed
neutrinos and right-handed antineutrinos have ever been detected \cite{PhysRev.109.1015,Agashe:2014kda}.

One open problem in neutrino physics is the determination of their
nature. Because they are electrically neutral two possibilities exist
\cite{Mohapatra:724618,Zralek:1997sa,fukugita2003physics,0034-4885-67-2-R01,giunti2007fundamentals,Petcov:2013poa}:
either neutrinos are distinct from their antiparticles and hence of
the Dirac type (implying lepton number conservation), or they are
equal to their antiparticles and hence of the Majorana type (implying
lepton number violation). Theoretically, the latter is the preferred
alternative for two main reasons: it provides a natural explanation
for light neutrino masses through the see-saw mechanism \cite{0034-4885-67-2-R01},
and allows for models of baryogenesis from leptogenesis that could
explain the baryon asymmetry of the Universe \cite{fukugita2003physics,doi:10.1080/00107514.2012.701096,CBO9781107358362A095,doi:10.1142/S0217751X14300282}.
Thus, in most extensions to the standard model it is \emph{ab initio
}assumed that neutrinos are Majorana particles \cite{CBO9781107358362A095,Mohapatra:2005wg}. 

When calculating amplitudes of neutrino processes it is found that,
regardless of the process, the difference in matrix elements for Dirac
and Majorana types is proportional to the ratio of the neutrino mass
$m_{\nu}$ to its energy $E_{\nu}$: $m_{\nu}/E_{\nu}$, which becomes
negligible small for high-energetic neutrinos with small masses \cite{Mohapatra:724618}.
This result makes very difficult the experimental determination of
the nature of the neutrino, with the best option provided by neutrinoless
double beta decay experiments, a process that can only occur if neutrinos
are Majorana particles. The present experimental status in this regard
is inconclusive \cite{Agashe:2014kda}, but there are both ongoing
and planned experiments that could resolve the issue in a few years
\cite{Bilenky:2012qi,Maneschg2015188}.

In this letter I address the problem of the neutrino nature, from
the perspective of canonical quantum field theory, and show that it
must necessarily be of the Majorana type. This is done by assuming
that a general neutrino with definite mass is described by a massive
Dirac field operator and showing that, nonetheless, the physical one-particle
states must be identified with the antiparticles ones because they
can be rotated to each other, violating lepton number conservation
in the process. I will also show that no such rotations are possible
for massive Majorana field operators. The procedure involves momentum-dependent
transformations acting on free field operators in an analogous way
to the standard discrete transformations of parity and charge conjugation,
with no models beyond the standard model assumed or required. Natural
units with $\hbar=c=1$ and the Weyl representation of the gamma matrices

\begin{equation}
\gamma^{0}=\left(\begin{array}{rr}
0 & 1\\
1 & 0
\end{array}\right),\,\boldsymbol{\gamma}=\left(\begin{array}{rr}
0 & \boldsymbol{\sigma}\\
-\boldsymbol{\sigma} & 0
\end{array}\right),\label{eq:gamma matrices}
\end{equation}

\noindent here denoted in $2\times2$ block form, with $\boldsymbol{\sigma}=\left(\sigma^{1},\sigma^{2},\sigma^{3}\right)$
the standard Pauli matrices, will be used throughout.

\section{Dirac case}

Let $a_{\lambda}\left(\mathbf{p}\right)$ and $b_{\lambda}\left(\mathbf{p}\right)$
respectively denote annihilation operators of Dirac particles and
antiparticles of a given helicity $\lambda=\pm$, and $\hat{a}_{\pm}\left(\mathbf{p}\right)$
represent the corresponding operators for Majorana particles. In weak
processes such as the decay of charged pions, the neutrinos produced
can be scattered off a nuclear target resulting in a flux of muons.
Neutrinos coming from $\pi^{+}$ decay produce muons, while antineutrinos
coming from $\pi^{-}$ decay produce antimuons. Since neutrinos and
antineutrinos have opposite helicities it is not possible to experimentally
distinguish if the production of muons is due to lepton number conservation,
in which case neutrinos are Dirac particles and \cite{Zralek:1997sa,Czakon:1999ed}

\begin{equation}
a\left(\mathbf{p}\right)\neq b\left(\mathbf{p}\right)\label{eq:Dirac condition}
\end{equation}

\noindent must hold irrespective of the helicity, or whether it is
the helicity of the particles the responsible for the production,
in which case neutrinos are Majorana particles and we must have

\begin{equation}
\hat{a}_{+}\left(\mathbf{p}\right)\neq\hat{a}_{-}\left(\mathbf{p}\right).\label{eq:Majorana condition}
\end{equation}

\noindent The goal is to show that Eq. (\ref{eq:Majorana condition})
is the only possibility, even if Eq. (\ref{eq:Dirac condition}) is
originally assumed. To this end let us consider one neutrino type
of definite mass and assume it is a Dirac particle, described by a
massive Dirac field. In the helicity basis, the field expansion is
given by \cite{duncan2012conceptual}

\begin{multline}
\Psi(x)=\int\frac{d^{3}p}{\left(2\pi\right)^{3}}\frac{1}{\sqrt{2E_{\mathbf{p}}}}\sum_{\lambda=\pm}\left(u_{\lambda}\left(\mathbf{p}\right)a_{\lambda}\left(\mathbf{p}\right)e^{-ip.x}\right.\\
\left.+v_{\lambda}\left(\mathbf{p}\right)b_{\lambda}^{\dagger}\left(\mathbf{p}\right)e^{ip.x}\right),\label{eq:massiveDirac}
\end{multline}

\noindent with the usual equal time anti-commutation relations satisfied
by both field and operators \cite{peskin1995introduction}

\begin{gather}
\begin{gathered}\left\{ \Psi_{\alpha}\left(\mathbf{x}\right),\Psi_{\beta}^{\dagger}\left(\mathbf{y}\right)\right\} =\delta^{3}\left(\mathbf{x}-\mathbf{y}\right)\delta_{\alpha\beta},\\
\left\{ \Psi_{\alpha}\left(\mathbf{x}\right),\Psi_{\beta}\left(\mathbf{y}\right)\right\} =\left\{ \Psi_{\alpha}^{\dagger}\left(\mathbf{x}\right),\Psi_{\beta}^{\dagger}\left(\mathbf{y}\right)\right\} =0,\\
\left\{ a_{\lambda}\left(\mathbf{p}\right),a_{\lambda'}^{\dagger}\left(\mathbf{q}\right)\right\} =\left\{ b_{\lambda}\left(\mathbf{p}\right),b_{\lambda'}^{\dagger}\left(\mathbf{q}\right)\right\} =\left(2\pi\right)^{3}\delta^{3}\left(\mathbf{p}-\mathbf{q}\right)\delta_{\lambda\lambda'}.
\end{gathered}
\label{eq:field comm}
\end{gather}

\noindent The positive- and negative-energy spinors $u_{\lambda}\left(\mathbf{p}\right)$
and $v_{\lambda}\left(\mathbf{p}\right)$ are four-component Dirac
spinors in the helicity basis. Explicitly

\begin{align}
\begin{split}u_{\lambda}\left(\mathbf{p}\right)= & \begin{pmatrix}\sqrt{E-\lambda\left|\mathbf{p}\right|}\xi_{\lambda}\left(\mathbf{p}\right)\\
\sqrt{E+\lambda\left|\mathbf{p}\right|}\xi_{\lambda}\left(\mathbf{p}\right)
\end{pmatrix},\\
v_{\lambda}\left(\mathbf{p}\right)= & \begin{pmatrix}-\lambda\sqrt{E+\lambda\left|\mathbf{p}\right|}\xi_{-\lambda}\left(\mathbf{p}\right)\\
\lambda\sqrt{E-\lambda\left|\mathbf{p}\right|}\xi_{-\lambda}\left(\mathbf{p}\right)
\end{pmatrix},
\end{split}
\label{eq:bispinors}
\end{align}

\noindent with $\lambda=\pm$ and $\xi_{\lambda}$ two-component helicity
spinors. Taking the three-momentum in spherical polar coordinates
$\mathbf{\hat{p}}=\left(\sin\theta\cos\varphi,\sin\theta\sin\varphi,\cos\theta\right),$
solutions to the helicity eigenvalue equation $\boldsymbol{\sigma}\cdot\mathbf{\hat{p}}\,\xi_{\lambda}(\mathbf{p})=\lambda\xi_{\lambda}(\mathbf{p})$
are readily found

\begin{equation}
\begin{array}{cc}
\xi_{+}(\mathbf{p})=\begin{pmatrix}\cos\left(\frac{\theta}{2}\right)\\
e^{i\varphi}\sin\left(\frac{\theta}{2}\right)
\end{pmatrix}, & \xi_{-}(\mathbf{p})=\begin{pmatrix}-e^{-i\varphi}\sin\left(\frac{\theta}{2}\right)\\
\cos\left(\frac{\theta}{2}\right)
\end{pmatrix}.\end{array}\label{eq:spinors}
\end{equation}

\noindent Then it is straightforward to verify the following relations:
$\left(\slashed{p}-m\right)u_{\lambda}\left(\mathbf{p}\right)=0$,
$\left(\slashed{p}+m\right)v_{\lambda}\left(\mathbf{p}\right)=0$,
with $\slashed{p}\equiv\gamma^{\mu}p_{\mu}$, and

\begin{align}
\begin{split}\overline{u}_{\lambda}\left(\mathbf{p}\right)u_{\lambda'}\left(\mathbf{p}\right) & =2m\delta_{\lambda,\lambda'},\\
\overline{v}_{\lambda}\left(\mathbf{p}\right)v_{\lambda'}\left(\mathbf{p}\right) & =-2m\delta_{\lambda,\lambda'},\\
\overline{u}_{\lambda}\left(\mathbf{p}\right)v_{\lambda'}\left(\mathbf{p}\right) & =0,
\end{split}
\label{eq:orthonormal}
\end{align}

\noindent where the over bar represents the Dirac adjoint $\overline{u}\equiv u^{\dagger}\gamma^{0}$.
Positive- and negative-energy spinors are also related to one another
by charge conjugation

\begin{align}
\begin{split}i\gamma^{2}u_{\lambda}^{*}\left(\mathbf{p}\right) & =v_{\lambda}\left(\mathbf{p}\right),\\
i\gamma^{2}v_{\lambda}^{*}\left(\mathbf{p}\right) & =u_{\lambda}\left(\mathbf{p}\right).
\end{split}
\label{eq:spinors CC}
\end{align}

The one-particle states created by the field $\Psi$ in Eq. (\ref{eq:massiveDirac})
and its Hermitian conjugate are summarized in Table 1, from them the
following transformations are constructed

\begin{equation}
\begin{aligned}\begin{split}P & =\alpha_{1}\left|\mathbf{p},-\right\rangle \left\langle \mathbf{p},+\right|+\alpha_{2}\left|\mathbf{p},+\right\rangle \left\langle \mathbf{p},-\right|\\
 & +\alpha_{3}\left|\mathbf{\overline{p}},-\right\rangle \left\langle \mathbf{\overline{p}},+\right|+\alpha_{4}\left|\mathbf{\overline{p}},+\right\rangle \left\langle \mathbf{\overline{p}},-\right|,
\end{split}
\end{aligned}
\label{eq:P-trans}
\end{equation}

\begin{align}
\begin{split}C & =\beta_{1}\left|\mathbf{p},-\right\rangle \left\langle \overline{\mathbf{p}},-\right|+\beta_{2}\left|\mathbf{\overline{p}},-\right\rangle \left\langle \mathbf{p},-\right|\\
 & +\beta_{3}\left|\mathbf{p},+\right\rangle \left\langle \mathbf{\overline{p}},+\right|+\beta_{4}\left|\mathbf{\overline{p}},+\right\rangle \left\langle \mathbf{p},+\right|,
\end{split}
\label{eq:C-trans}
\end{align}

\noindent with $\alpha_{i},\beta_{i},\,\,i=1,\ldots,4$ complex coefficients
of unit modulus $\left|\alpha_{i}\right|^{2}=\left|\beta_{i}\right|^{2}=1,\,\,i=1,\ldots,4,$
to assure unitarity. Applied to the states these transformations yield

\begin{equation}
\begin{array}{cc}
\begin{split}P\left|\mathbf{p},-\right\rangle  & =\alpha_{2}\left|\mathbf{p},+\right\rangle ,\\
P\left|\mathbf{p},+\right\rangle  & =\alpha_{1}\left|\mathbf{p},-\right\rangle ,\\
P\left|\mathbf{\overline{p}},-\right\rangle  & =\alpha_{4}\left|\overline{\mathbf{p}},+\right\rangle ,\\
P\left|\mathbf{\overline{p}},+\right\rangle  & =\alpha_{3}\left|\mathbf{\overline{p}},-\right\rangle ,
\end{split}
 & \begin{split}C\left|\mathbf{p},-\right\rangle  & =\beta_{2}\left|\mathbf{\overline{p}},-\right\rangle ,\\
C\left|\mathbf{p},+\right\rangle  & =\beta_{4}\left|\mathbf{\overline{p}},+\right\rangle ,\\
C\left|\mathbf{\overline{p}},-\right\rangle  & =\beta_{1}\left|\mathbf{p},-\right\rangle ,\\
C\left|\mathbf{\overline{p}},+\right\rangle  & =\beta_{3}\left|\mathbf{p},+\right\rangle .
\end{split}
\end{array}\label{eq:P and  C action}
\end{equation}

\begin{table}
\begin{centering}
\begin{tabular}{|c|c|c|c|}
\hline 
State  & Definition  & Lepton number  & Particle\tabularnewline
\hline 
\hline 
$\left|\mathbf{p},-\right\rangle $  & $a_{-}^{\dagger}\left(\mathbf{p}\right)\left|0\right\rangle $  & 1  & LH neutrino\tabularnewline
\hline 
$\left|\overline{\mathbf{p}},+\right\rangle $  & $b_{+}^{\dagger}\left(\mathbf{p}\right)\left|0\right\rangle $  & -1  & RH antineutrino\tabularnewline
\hline 
$\left|\mathbf{p},+\right\rangle $  & $a_{+}^{\dagger}\left(\mathbf{p}\right)\left|0\right\rangle $  & 1  & RH neutrino\tabularnewline
\hline 
$\left|\mathbf{\overline{p}},-\right\rangle $  & $b_{-}^{\dagger}\left(\mathbf{p}\right)\left|0\right\rangle $  & -1  & LH antineutrino\tabularnewline
\hline 
\end{tabular}
\par\end{centering}

\caption{One particle states created from the vacuum by $\Psi$ and $\Psi^{\dagger}$.
The states are assumed orthonormal and a possible normalization factor
in the second column has been omitted. For neutrinos the $U(1)$ charge
is associated with lepton number and not with electrical charge. In
the case of states the over bar represents the antiparticle, and the
terms left-handed (LH) and right-handed (RH) refer to $\mp$ helicity,
respectively.}
\end{table}

In the one-particle states basis, with rows $\left\{ \left|\mathbf{p},-\right\rangle ,\left|\mathbf{p},+\right\rangle ,\left|\mathbf{\overline{p}},-\right\rangle ,\left|\mathbf{\overline{p}},+\right\rangle \right\} $
and the corresponding bras as columns, explicit matrix representations
of Eqs. (\ref{eq:P-trans}) and (\ref{eq:C-trans}) are given by

\begin{equation}
\begin{array}{cc}
P=\begin{pmatrix}0 & \alpha_{1} & 0 & 0\\
\alpha_{2} & 0 & 0 & 0\\
0 & 0 & 0 & \alpha_{3}\\
0 & 0 & \alpha_{4} & 0
\end{pmatrix}, & C=\begin{pmatrix}0 & 0 & \beta_{1} & 0\\
0 & 0 & 0 & \beta_{3}\\
\beta_{2} & 0 & 0 & 0\\
0 & \beta_{4} & 0 & 0
\end{pmatrix}.\end{array}\label{eq:P and C matrices}
\end{equation}

\noindent Then the following properties are easily verified

\begin{gather}
\begin{gathered}\det P=\alpha_{1}\alpha_{2}\alpha_{3}\alpha_{4},\\
\det C=\beta_{1}\beta_{2}\beta_{3}\beta_{4},\\
PP^{\dagger}=P^{\dagger}P=CC^{\dagger}=C^{\dagger}C=\mathbbm{1}_{4}.
\end{gathered}
\label{eq:PC properties}
\end{gather}

\noindent Thus, the $P$ and $C$ transformations can be either rotations
or reflections (respectively of $\pm1$ determinant), depending on
the choice of the $\alpha_{i}$ and $\beta_{i}$ phases. Assuming
invariance of the vacuum under the transformations: $P,C\left|0\right\rangle =\left|0\right\rangle $,
Eq. (\ref{eq:P and  C action}) yields

\begin{equation}
\begin{array}{cc}
\begin{split}Pa_{-}^{\dagger}\left(\mathbf{p}\right)P^{\dagger} & =\alpha_{2}a_{+}^{\dagger}\left(\mathbf{p}\right),\\
Pa_{+}^{\dagger}\left(\mathbf{p}\right)P^{\dagger} & =\alpha_{1}a_{-}^{\dagger}\left(\mathbf{p}\right),\\
Pb_{-}^{\dagger}\left(\mathbf{p}\right)P^{\dagger} & =\alpha_{4}b_{+}^{\dagger}\left(\mathbf{p}\right),\\
Pb_{+}^{\dagger}\left(\mathbf{p}\right)P^{\dagger} & =\alpha_{3}b_{-}^{\dagger}\left(\mathbf{p}\right),
\end{split}
 & \begin{split}Ca_{-}^{\dagger}\left(\mathbf{p}\right)C^{\dagger} & =\beta_{2}b_{-}^{\dagger}\left(\mathbf{p}\right),\\
Ca_{+}^{\dagger}\left(\mathbf{p}\right)C^{\dagger} & =\beta_{4}b_{+}^{\dagger}\left(\mathbf{p}\right),\\
Cb_{-}^{\dagger}\left(\mathbf{p}\right)C^{\dagger} & =\beta_{1}a_{-}^{\dagger}\left(\mathbf{p}\right),\\
Cb_{+}^{\dagger}\left(\mathbf{p}\right)C^{\dagger} & =\beta_{3}a_{+}^{\dagger}\left(\mathbf{p}\right).
\end{split}
\end{array}\label{eq:ops P and C}
\end{equation}

\noindent Up to phases, the $C$ transformation for operators in Eq.
(\ref{eq:ops P and C}) coincide with the standard charge conjugation
transformation $\mathcal{C}$ \cite{peskin1995introduction}, since
it exchanges particle and antiparticle without changing the helicity.
But the $P$ transformation does not correspond to the standard parity
transformation $\mathcal{P}$, even though it flips the helicity,
because the latter requires $\mathcal{P}a_{\lambda}\left(\mathbf{p}\right)\mathcal{P}^{\dagger}=a_{-\lambda}\left(-\mathbf{p}\right)$. 

Equations (\ref{eq:massiveDirac}) and (\ref{eq:ops P and C}) lead
to

\begin{multline}
P\Psi(x)P^{\dagger}=\int\frac{d^{3}p}{\left(2\pi\right)^{3}}\frac{1}{\sqrt{2E_{\mathbf{p}}}}\left(\alpha_{1}^{*}u_{+}\left(\mathbf{p}\right)a_{-}\left(\mathbf{p}\right)e^{-ip.x}\right.\\
+\alpha_{2}^{*}u_{-}\left(\mathbf{p}\right)a_{+}\left(\mathbf{p}\right)e^{-ip.x}+\alpha_{3}v_{+}\left(\mathbf{p}\right)b_{-}^{\dagger}\left(\mathbf{p}\right)e^{ip.x}\\
\left.+\alpha_{4}v_{-}\left(\mathbf{p}\right)b_{+}^{\dagger}\left(\mathbf{p}\right)e^{ip.x}\right),\label{eq:P on field}
\end{multline}

\noindent and, in order for this to represent a transformation of
the field, we need to make a consistent choice for the $\alpha_{i}$
coefficients and flip the helicity of the spinors using some unitary
matrix. The latter can be achieved by momentum-dependent transformations,
built from the spinors themselves in analogy with Eq. (\ref{eq:P-trans}).
One such transformation is given by

\begin{multline}
R(\mathbf{p})=-i\cos\left(\varphi\right)\sin\left(\theta\right)\gamma^{0}\gamma^{1}\gamma^{2}\\
+i\sin^{2}\left(\frac{\theta}{2}\right)\sin\left(2\varphi\right)\gamma^{0}\gamma^{1}\gamma^{3}\\
+i\left(\cos^{2}\left(\frac{\theta}{2}\right)-2\cos\left(2\varphi\right)\sin^{2}\left(\frac{\theta}{2}\right)\right)\gamma^{0}\gamma^{2}\gamma^{3},\label{eq:RotP}
\end{multline}

\noindent which is unitary and of unit determinant, hence a rotation.
In terms of the spinors it reads

\begin{align}
\begin{split}R(\mathbf{p}) & =\frac{1}{2m}\left(u_{+}\left(\mathbf{p}\right)\overline{u}_{-}\left(\mathbf{p}\right)+u_{-}\left(\mathbf{p}\right)\overline{u}_{+}\left(\mathbf{p}\right)\right.\\
 & \left.-v_{+}\left(\mathbf{p}\right)\overline{v}_{-}\left(\mathbf{p}\right)-v_{-}\left(\mathbf{p}\right)\overline{v}_{+}\left(\mathbf{p}\right)\right),
\end{split}
\label{eq:P matrix decomp}
\end{align}

\noindent from which, together with Eq. (\ref{eq:orthonormal}), it
is directly verified that

\begin{align}
\begin{split}R\left(\mathbf{p}\right)u_{\lambda}\left(\mathbf{p}\right) & =u_{-\lambda}\left(\mathbf{p}\right),\\
R\left(\mathbf{p}\right)v_{\lambda}\left(\mathbf{p}\right) & =v_{-\lambda}\left(\mathbf{p}\right).
\end{split}
\label{eq:RotP spinors}
\end{align}

\noindent Thus, Eqs. (\ref{eq:P on field}) and (\ref{eq:RotP spinors})
lead to

\begin{equation}
P\Psi(x)P^{\dagger}=R\left(\mathbf{p}\right)\Psi(x),\label{eq:P compl alt}
\end{equation}

\noindent provided that

\begin{equation}
\alpha_{i}=1,\,\,i=1,\ldots,4,\label{eq:P-phase alt}
\end{equation}

\noindent which in turn, from Eq. (\ref{eq:PC properties}), makes
$P$ a rotation. If the $\hat{\mathbf{z}}$ axis is taken as the quantization
axis, as is commonly chosen, we can take $\theta=\varphi=0$ and then
the spinors in Eq. (\ref{eq:spinors}) reduce to eigenstates of $\sigma^{3}$
and, from Eq. (\ref{eq:RotP}), the matrix $R(\mathbf{p})$ reduces
to $i\gamma^{0}\gamma^{2}\gamma^{3}$.

Let us now turn to the $C$ transformation, Eqs. (\ref{eq:massiveDirac})
and (\ref{eq:ops P and C}) yield

\begin{multline}
C\Psi(x)C^{\dagger}=\int\frac{d^{3}p}{\left(2\pi\right)^{3}}\frac{1}{\sqrt{2E_{\mathbf{p}}}}\left(\beta_{4}^{*}u_{+}\left(\mathbf{p}\right)b_{+}\left(\mathbf{p}\right)e^{-ip.x}\right.\\
+\beta_{2}^{*}u_{-}\left(\mathbf{p}\right)b_{-}\left(\mathbf{p}\right)e^{-ip.x}+\beta_{3}v_{+}\left(\mathbf{p}\right)a_{+}^{\dagger}\left(\mathbf{p}\right)e^{ip.x}\\
\left.+\beta_{1}v_{-}\left(\mathbf{p}\right)a_{-}^{\dagger}\left(\mathbf{p}\right)e^{ip.x}\right).\label{eq:C on field}
\end{multline}

\noindent Then, using Eq. (\ref{eq:spinors CC}) and choosing the
phases

\begin{equation}
\beta_{i}=1,\,\,i=1,\ldots,4,\label{eq:C-phase}
\end{equation}

\noindent the field transformation is

\begin{equation}
C\Psi(x)C^{\dagger}=i\gamma^{2}\Psi^{*}(x).\label{eq:C trans compl}
\end{equation}

\noindent The matrix $i\gamma^{2}$ is a rotation, and from Eqs. (\ref{eq:PC properties})
and (\ref{eq:C-phase}) we get that $C$ is also a rotation. Furthermore,
its action on the field is lineal, even though the complex conjugate
field appears in Eq. (\ref{eq:C trans compl}). This implies that
charge conjugation can be consistently implemented as a rotation,
just as in the case of the $P$ transformation.

Already from Eqs. (\ref{eq:PC properties}), (\ref{eq:ops P and C}),
and (\ref{eq:C-phase}) we get that particles and antiparticles of
the same helicity can be rotated to each other and so they must be
identified $a_{\lambda}\left(\mathbf{p}\right)=b_{\lambda}\left(\mathbf{p}\right)$,
in contradiction with Eq. (\ref{eq:Dirac condition}). However, weak
interactions involve left-handed neutrinos and right-handed antineutrinos,
meaning we really ought to compare $a$'s and $b$'s of different
helicities. But first let us check the commutation relations between
the $C$ transformation and the Hamiltonian and lepton number operators.
Following the usual procedure \cite{peskin1995introduction} and using
Eqs. (\ref{eq:massiveDirac}) and (\ref{eq:field comm}) we obtain

\begin{align}
\begin{split}\mathcal{H} & =\int d^{3}x\overline{\Psi}\left(-i\boldsymbol{\gamma}\cdot\nabla+m\right)\Psi\\
 & =\int\frac{d^{3}p}{\left(2\pi\right)^{3}}\sum_{\lambda=\pm}E_{\mathbf{p}}\left(a_{\lambda}^{\dagger}\left(\mathbf{p}\right)a_{\lambda}\left(\mathbf{p}\right)+b_{\lambda}^{\dagger}\left(\mathbf{p}\right)b_{\lambda}\left(\mathbf{p}\right)\right),
\end{split}
\label{eq:Hamiltonian}
\end{align}

\begin{align}
\begin{split}L & =\int d^{3}x\Psi^{\dagger}(x)\Psi(x)\\
 & =\int\frac{d^{3}p}{\left(2\pi\right)^{3}}\sum_{\lambda=\pm}\left(a_{\lambda}^{\dagger}\left(\mathbf{p}\right)a_{\lambda}\left(\mathbf{p}\right)-b_{\lambda}^{\dagger}\left(\mathbf{p}\right)b_{\lambda}\left(\mathbf{p}\right)\right).
\end{split}
\label{eq:charge op}
\end{align}

\noindent The operator in Eq. (\ref{eq:charge op}) is the charge
operator for charged fermions, but for neutrinos it acts as a lepton
number operator $L$ \cite{giunti2007fundamentals}. Then it follows
from Eqs. (\ref{eq:ops P and C}), (\ref{eq:Hamiltonian}), and (\ref{eq:charge op})
that

\noindent 
\begin{equation}
\left[\mathcal{H},L\right]=\left[\mathcal{H},C\right]=\left\{ L,C\right\} =0,\label{eq:comm relations}
\end{equation}

\noindent where the square and curly brackets respectively represent
the commutator and the anti-commutator. These relations imply that
the physical states can either be eigenstates of $\mathcal{H}$ and
$L$ or of $\mathcal{H}$ and $C$. But the action of the $C$ rotation
means that lepton number conservation is violated even if the former
alternative is chosen, leading to neutrino states effectively behaving
as Majorana particles.

Let us now consider the combined action of the $P$ and $C$ transformations.
Using the phase conventions of Eqs. (\ref{eq:P-phase alt}) and (\ref{eq:C-phase})
we get from Eqs. (\ref{eq:PC properties}) and (\ref{eq:ops P and C})
that $CP$ is a rotation that changes a particle into the antiparticle
of opposite helicity, and from Eqs. (\ref{eq:spinors CC}) and (\ref{eq:RotP spinors})
the field transformation is

\begin{equation}
CP\Psi(x)\left(CP\right)^{\dagger}=i\gamma^{2}R^{*}\left(\mathbf{p}\right)\Psi^{*}(x),\label{eq:CP field}
\end{equation}

\noindent and it can be verified that the combined $CP$ transformation
also satisfies Eq. (\ref{eq:field comm}). The $CP$ transformation
implies that opposite helicity particles and antiparticles can be
rotated to each other, and so they must be identified, again in contradiction
with Eq. (\ref{eq:Dirac condition}). Thus, we have proved that Eq.
(\ref{eq:Dirac condition}) is untenable in general, leaving Eq. (\ref{eq:Majorana condition})
as the only possibility, and it remains to show that this is indeed
the case.

\section{Majorana case}

The Majorana field expansion is obtained directly from Eq. (\ref{eq:massiveDirac}),
with $a_{\lambda}\left(\mathbf{p}\right)=b_{\lambda}\left(\mathbf{p}\right)\equiv\hat{a}_{\lambda}\left(\mathbf{p}\right)$

\begin{multline}
\nu(x)=\int\frac{d^{3}p}{\left(2\pi\right)^{3}}\frac{1}{\sqrt{2E_{\mathbf{p}}}}\sum_{\lambda=\pm}\left(u_{\lambda}\left(\mathbf{p}\right)\hat{a}_{\lambda}\left(\mathbf{p}\right)e^{-ip.x}\right.\\
\left.+\zeta v_{\lambda}\left(\mathbf{p}\right)\hat{a}_{\lambda}^{\dagger}\left(\mathbf{p}\right)e^{ip.x}\right)\label{eq:massiveMajorana}
\end{multline}

\noindent where, for generality sake, a creation phase $\zeta$ has
been introduced \cite{Mohapatra:724618}. The field satisfies the
Majorana condition $\nu(x)=i\zeta^{*}\gamma^{2}\nu^{*}(x)$. Since
now there are only two one-particle states available, the ones created
off the vacuum by $\hat{a}_{+}^{\dagger}\left(\mathbf{p}\right)$
and $\hat{a}_{-}^{\dagger}\left(\mathbf{p}\right)$, there is only
one type of transformation connecting states of opposite helicity.
Denoting it by $\Omega$, it is given by

\begin{equation}
\Omega=\delta_{1}\left|\mathbf{p},-\right\rangle \left\langle \mathbf{p},+\right|+\delta_{2}\left|\mathbf{p},+\right\rangle \left\langle \mathbf{p},-\right|=\begin{pmatrix}0 & \delta_{2}\\
\delta_{1} & 0
\end{pmatrix},\label{eq:CP Majorana}
\end{equation}

\noindent where $\delta_{1,2}$ are phases, the states are now $\left|\mathbf{p},\pm\right\rangle =\hat{a}_{\pm}^{\dagger}\left(\mathbf{p}\right)\left|0\right\rangle $,
and the matrix form in the last equality is obtained in this basis.
It follows that

\begin{equation}
\begin{array}{cc}
\Omega\Omega^{\dagger}=\Omega^{\dagger}\Omega=\mathbbm{1}_{2}, & \det\Omega=-\delta_{1}\delta_{2}.\end{array}\label{eq:CP Maj. det}
\end{equation}

\noindent In order for $\Omega$ to be a rotation we must have $\delta_{1}=-1/\delta_{2}$,
and so we can generally write

\begin{equation}
\delta_{1}=-\delta_{2}^{*}=e^{i\omega},\label{eq:Majorana phase}
\end{equation}

\noindent with $\omega$ a real parameter. For the operators, and
again assuming invariance of the vacuum, Eqs. (\ref{eq:CP Majorana})
and (\ref{eq:Majorana phase}) imply

\begin{equation}
\begin{array}{cc}
\Omega\hat{a}_{+}^{\dagger}\left(\mathbf{p}\right)\Omega^{\dagger}=e^{i\omega}\hat{a}_{-}^{\dagger}\left(\mathbf{p}\right), & \Omega\hat{a}_{-}^{\dagger}\left(\mathbf{p}\right)\Omega^{\dagger}=-e^{-i\omega}\hat{a}_{+}^{\dagger}\left(\mathbf{p}\right).\end{array}\label{eq:CP majorana ops}
\end{equation}

\noindent which yields, upon applying it to the Majorana field in
Eq. (\ref{eq:massiveMajorana})

\begin{multline}
\Omega\nu(x)\Omega^{\dagger}=\int\frac{d^{3}p}{\left(2\pi\right)^{3}}\frac{1}{\sqrt{2E_{\mathbf{p}}}}\left(e^{-i\omega}u_{+}\left(\mathbf{p}\right)\hat{a}_{-}\left(\mathbf{p}\right)e^{-ip.x}\right.\\
-e^{i\omega}u_{-}\left(\mathbf{p}\right)\hat{a}_{+}\left(\mathbf{p}\right)e^{-ip.x}+\zeta e^{i\omega}v_{+}\left(\mathbf{p}\right)\hat{a}_{-}^{\dagger}\left(\mathbf{p}\right)e^{ip.x}\\
\left.-\zeta e^{-i\omega}v_{-}\left(\mathbf{p}\right)\hat{a}_{+}^{\dagger}\left(\mathbf{p}\right)e^{ip.x}\right).\label{eq:CP Majorana field}
\end{multline}

\noindent This last result is analogous to the one in Eq. (\ref{eq:P on field}),
therefore we need a unitary matrix $\hat{R}$ acting on the spinors,
as in Eq. (\ref{eq:RotP spinors}), that flips the helicity and provides
the required phases. Allowing for arbitrary phases in the expansion
of Eq. (\ref{eq:P matrix decomp}) provides the most general transformation
between different helicity spinors of the same energy. Thus, a suitable
modification of Eq. (\ref{eq:P matrix decomp}) accomplishes the required
transformation, this is

\begin{align}
\begin{split}\hat{R}(\mathbf{p}) & =\frac{1}{2m}\left(e^{-i\omega}u_{+}\left(\mathbf{p}\right)\overline{u}_{-}\left(\mathbf{p}\right)-e^{i\omega}u_{-}\left(\mathbf{p}\right)\overline{u}_{+}\left(\mathbf{p}\right)\right.\\
 & \left.-e^{i\omega}v_{+}\left(\mathbf{p}\right)\overline{v}_{-}\left(\mathbf{p}\right)+e^{-i\omega}v_{-}\left(\mathbf{p}\right)\overline{v}_{+}\left(\mathbf{p}\right)\right),
\end{split}
\label{eq:P non-unitary}
\end{align}

\noindent but at the cost of loosing unitarity, since now $\hat{R}^{\dagger}(\mathbf{p})\hat{R}(\mathbf{p})\neq\mathbbm{1}_{4}$
as is readily verified. This result shows that it is not possible
to consistently make $\Omega$ a rotation of the Majorana field, which
implies that the operators $\hat{a}_{\lambda}\left(\mathbf{p}\right)$
cannot be rotated to each other, and so Eq. (\ref{eq:Majorana condition})
must hold. The task is now complete and it is concluded that the neutrino
is a Majorana particle.

\section{Concluding remarks}

The arguments presented in this letter, showing the Majorana nature
of the neutrino, are ultimately related to the degrees of freedom
of Dirac and Majorana particles. It is the fact that a Dirac particle
possesses four degrees of freedom which allows for the existence of
the $C$ and $P$ transformations, as expressed in Eqs. (\ref{eq:P-trans})
and (\ref{eq:C-trans}), which lead to the identification of neutrino
and antineutrino states, in direct contradiction to the fact that
they should be different, and to lepton number violation. It is also
the fact that a Majorana particle has only two degrees of freedom
which prevents a similar contradiction to arise if the neutrino is
assumed to be a Majorana particle, leaving such an assumption as the
only viable possibility. These considerations also show that the conclusion
reached is fundamental, and must hold independently of the mass generation,
oscillation, and mixing mechanisms, all of which cannot alter the
fundamental degrees of freedom. The final vindication of the Majorana
nature of the neutrino must come from the experimental confirmation
of neutrinoless double beta decays.

As a final note, Eqs. (\ref{eq:C trans compl}) and (\ref{eq:CP field})
also imply that charge conservation is violated for free charged fermions
or, equivalently, that free fermions are essentially Majorana particles,
irrespective of their charge. This, of course, no longer holds for
interactive charged fermions, since the operators $C$ and $CP$ do
not commute with the current operator $\hat{J}^{\mu}$ defined in
the usual way \cite{peskin1995introduction}. These results could
be relevant in explaining the Universe baryon asymmetry, a possibility
that calls for further investigation.

\bibliographystyle{apsrev4-1}

\end{document}